\documentclass[twocolumn]{aastex63}

\usepackage{natbib}
\usepackage{color}
\usepackage{mathtools}
\usepackage{hyperref} 

\def    \apjl  		{\rm {ApJL}}

\def    \apjl  		{\rm {ApJL}}

\def	\cm		{\,{\rm {cm}}}
\def	\K		{\,{\rm K}}
\def	\g		{\,{\rm {g}}}
\def	\mum	{\,{\mu \rm{m}}}

\def \bea {\begin{eqnarray}}
\def \ena {\end{eqnarray}}

\def    \bB     {\bf  B}

\def	\bB	{{\bf B}}
\def	\B	{{\rm B}}

\def 	\bE	{{\bf E}}

\def	\bJ	{{\bf J}}

\def    \bmu    {{\hbox{\boldsym\char'026}}}	

\def    \bomega {{\hbox{\boldsym\char'041}}}	
\def	\bp	{{\bf p}}

\def	\C	{{\rm C}}

\def	\cm	{\,{\rm cm}}

\def	\max	{\,{\rm max}}
\def	\d	{{\rm d}}

\def	\D	{{\rm D}}

\def	\erg	{\,{\rm erg}}

\def	\g	{\,{\rm g}}
\def	\gas	{\,{\rm gas}}
\def	\G	{{\rm G}}

\def	\H	{{\rm H}}

\def	\N	{{\rm N}}

\def	\s	{\,{\rm s}}

\def	\AU	{\,{\rm au}}

\def	\V	{{\rm V}}
\def	\Bar	{{\rm Bar}}
\def	\rad	{{\rm rad}}
\def	\yr	{{\rm yr}}

\def    \kv     	{\bf  k}

\def    \gas     	{{\rm gas}}

\font\mib=cmmib10

\def\bomega{\hbox{\mib\char"21}}

\def\bmu{\hbox{\mib\char"16}}

\begin{document}
\shorttitle{Grain alignment and disruption in exoplanet atmospheres}
\shortauthors{Thiem Hoang and A. Lazarian}
\title{Grain alignment and rotational disruption by radiative torques in exoplanet atmospheres}

\author{Thiem Hoang}
\affiliation{Korea Astronomy and Space Science Institute, Daejeon 34055, Republic of Korea, \href{mailto:thiemhoang@kasi.re.kr}{thiemhoang@kasi.re.kr}}
\affiliation{Korea University of Science and Technology, 217 Gajeong-ro, Yuseong-gu, Daejeon, 34113, Republic of Korea}

\author{Alex Lazarian}
\affiliation{Department of Astronomy, University of Wisconsin, 475 North Charter Street, Madison, WI 53706, USA; Center for Computation Astrophysics, Flatiron Institute, 162 5th Avenue, New York, NY 10010, USA}

\begin{abstract}
Dust clouds are ubiquitous in the atmospheres of hot Jupiters and affect their observable properties. The alignment of dust grains in the clouds and resulting dust polarization is a promising method to study magnetic fields of exoplanets. Moreover, the grain size distribution plays an important role in physical and chemical processes in the atmospheres, which is rather uncertain in atmospheres. In this paper, we first study grain alignment of dust grains in the atmospheres of hot Jupiters by RAdiative Torques (RATs). We find that silicate grains can be aligned by RATs with the magnetic fields (B-RAT) due to strong magnetic fields of hot Jupiters, but carbonaceous grains of diamagnetic material tend to be aligned with the radiation direction (k-RAT). At a low altitude of $r<2R_{\rm p}$ with $R_{\rm p}$ planet radius, only large grains can be aligned, but tiny grains of $a\sim 0.01\mum$ can be aligned at a high altitude of $r>3R_{\rm p}$. We then study rotational disruption of dust grains by the RAdiative Torque Disruption (RATD) mechanism. We find that large grains can be disrupted by RATD into smaller sizes. Grains of high tensile strength are disrupted at an altitude of $r>3R_{\rm p}$, but weak grains can be disrupted at a lower altitude. We suggest that the disruption of large grains into smaller ones can facilitate dust clouds to escape to high altitudes due to lower gravity and may explain the presence of high-altitude clouds in hot Jupiter as well as super-puff atmospheres.
 
\end{abstract}
\keywords{ISM: dust-extinction, ISM: general, radiation: dynamics, polarization, magnetic fields}

\section{Introduction}
Characterization of exoplanet atmospheres is a hot topic in astrophysics (\citealt{2010ARA&A..48..631S}; \citealt{Madhusudhan:2019gl}). Transmission spectroscopy of transiting exoplanets provides crucial information on chemical and physical properties of exoplanet atmospheres. The flat transmission spectra observed toward transiting exoplanets suggest the ubiquity of dust clouds at high altitudes in atmospheres that obscure part of the atmosphere and smooth out water absorption features (\citealt{Sing:2015hi}; \citealt{Barstow.2016}; see \citealt{Gao.2021} for reviews on exoplanet atmospheres) and \cite{Fortney.2021} for a review on hot Jupiters. 

The detection of clouds in hot Jupiter atmospheres was first reported in the hot Jupiter HD189733b with the Hubble Space Telescope (\citealt{2008MNRAS.385..109P}; \citealt{Sing:2009gw}); \citealt{{Pont:2008dc},{Pont:2013gw}}) and in hot Jupiter Kepler-7b by \cite{2013ApJ...776L..25D}.
Later on, more clouds are detected in the super-Earth exoplanets (\citealt{Kreidberg:2014gk}) and GJ436b (\citealt{2014Natur.505...66K}).
Clouds are also detected in {\it super-puffs} as revealed by featureless transmission spectra (e.g., \citealt{Chachan.2020}).

Dust formation in the atmospheres of hot Jupiters is expected to be analogous to dust formation in the atmosphere of brown dwarfs due to their similarity in the physical conditions. Dust formation due to condensation of gas in the atmospheres of brown dwarfs were studied in detail (\citealt{2001ApJ...556..872A}; \citealt{Fortney:2005dl}; \citealt{Helling:2006iy}). Following this scenario, dust formation primarily occurs in the dense region of low altitude. As a result, dust clouds are expected to be in low altitudes of exoplanet atmospheres (see \citealt{Fortney.2021} for a recent review on hot Jupiters). However, observations reveal the presence of clouds at high-altitudes (e.g., \citealt{Huitson:2012jh}; \citealt{Sing:2015hi}; \citealt{Barstow.2016}). It is suggested that dust clouds can be transported to high altitudes due to turbulence can mix dust (turbulent mixing) and bring small grains to the upper atmosphere. Micron-sized particles can be lofted by atmospheric circulation \citep{2013ApJ...777..100H}.

Several processes have been proposed to explain the presence of clouds at high-altitude of super-puffs, including dusty outflows (\citealt{j82}), dust formation by photochemical haze (\citealt{Gao:2020dq}; \citealt{Ohno.2021}).


Magnetic fields play a crucial role in exoplanet atmospheres, but their research is still unexplored. Magnetic fields can shield stellar winds and cosmic rays, which is an important parameter for habitability. Polarization of starlight as well as polarized thermal dust emission is a popular method to observe interstellar magnetic fields (\citealt{2012ARA&A..50...29C}), which is particularly promising for hot Jupiter and giant gas planets. Polarization in thermal emission from giant planets is discussed in \cite{Marley:2011iy} (see \citealt{2013arXiv1301.5627M} for a review). Recent study by \cite{Stark:2021} suggested that if grains are aligned, they can produce a polarization degree of orders of $\sim 1\%$. Following the modern understanding of grain alignment theory, dust grains in the hot Jupiter's atmospheres subject to stellar radiation can be efficiently aligned by RAdiative Torques (RATs; \citealt{2007MNRAS.378..910L}; \citealt{{Hoang:2008gb},{2016ApJ...831..159H}}). Therefore, the first goal of this paper is to study the possibility of tracing magnetic fields in exoplanet atmospheres using dust polarization using the leading RAT alignment theory (\citealt{LAH15}; \citealt{2015ARA&A..53..501A}).

Dust clouds play an essential role in the atmosphere, but the basic properties of dust (size distribution, shape, internal structure) are uncertain (see \citealt{2013arXiv1301.5627M} and \citealt{Fortney.2021} for reviews). 
In particular, the grain size distribution is the most important parameter controlling the physical and chemical processes in the atmosphere because the size distribution determines the wavelength-dependence absorption of starlight, which affects all observable properties of hot Jupiters (see \citealt{Fortney.2021}). It is also important for photoelectric heating effect to gas and thermal escape of atmosphere (\citealt{2019AREPS..47...67O}). However, the grain size distribution is poorly constrained. 

Previous studies consider the formation and coagulation of dust (e.g., \citealt{Lavvas.2017}; \citealt{Ohno:2020bj}). Growth of particles due to grain collisions is studied in \cite{Morley.2012} for brown dwarf atmosphere and exoplanet atmosphere \citep{Lavvas.2017} and the maximum grain size is found to be $1\mum$. Simulations of atmospheric escape including photoelectric heating is presented in \cite{Mitani:2020wa} where heating efficiency is adopted from the ISM. Size distribution of dust in atmosphere is found log-normal (\citealt{Morley:2015es}; \citealt{Nikolov:2014et}. Various models of the grain size distributions are available (\citealt{Lavvas.2017}; \citealt{Lacy.2020}), and the grain size distributions falls rapidly for micron-sized grains of $a>1-5\mum$ (see Figure 8 in \citealt{Gao.2021} for a review). \citealt{Ohno:2020bj} studied grain growth of fluffy structures and suggested that large aggregates of $\sim 10\mum$ can form. However, due to its proximity to the host star, dust grains are irradiated by intense radiation, and large grains can be disrupted rotationally by the Radiative Torque Disruption (RATD) mechanism (\citealt{Hoang:2019da}; \citealt{2019ApJ...876...13H}; see \citealt{2020Galax...8...52H} for a review). In this paper, we use the new RATD effect to constrain the upper cutoff of the grain size distribution. Previously, rotational disruption of grains by RATD in the atmosphere and intermediate layer of protoplanetary disks (PPDs) is presented in \cite{Tung:2020ew}.

The structure of the present paper is as follows. In Section \ref{sec:model}, we first describe the density and temperature profiles adopted for hot Jupiter atmospheres. In Section \ref{sec:RATs} we briefly review RATs for irregular grains and present analytical calculations of the RAT efficiency averaged over a stellar radiation field. Section \ref{sec:RATA} and \ref{sec:RATD} are devoted for study of grain alignment and rotational disruption by RATs, respectively. Discussion and summary of our main results are presented in Section \ref{sec:discuss} and \ref{sec:summary}.

\section{Physical model of hot Jupiter atmosphere}\label{sec:model}
\subsection{Gas density and temperature}

Let $R_{\rm p}$ and $M_{\rm p}$ be the radius and mass of the planet. We consider the one-dimensional model of the hot Jupiter's atmosphere as described in \cite{2009ApJ...693...23M} (see also \citealt{Fortney:2005dl}; \citealt{2008ApJ...678.1419F}) where one can obtain the gas density by numerically solving the hydrodynamic equation (\citealt{2009ApJ...693...23M}; \citealt{2019MNRAS.490.3760A}).

For simplicity, we assume the isothermal model for the atmosphere (see Appendix \ref{sec:apdx}. Thus, the gas number density decreases with the radius $r$ as follows
\bea
n_{\H}(r)=n_{p}\exp\left[\frac{GM_{\rm p}}{c_{s}^{2}} \left(\frac{1}{r}-\frac{1}{R_{\rm p}}\right)\right],
\ena
where $n_{p}$ is the gas number density at the planet surface with $r=R_{\rm p}$, and $c_{s}=(kT_{\gas}/m_{\H}\mu)^{1/2}$ with molecular weight $\mu$ is the sound speed (\citealt{2009ApJ...693...23M}; \citealt{2019MNRAS.490.3760A}; \citealt{Mitani:2020wa}).

\begin{figure}
\includegraphics[width=0.5\textwidth]{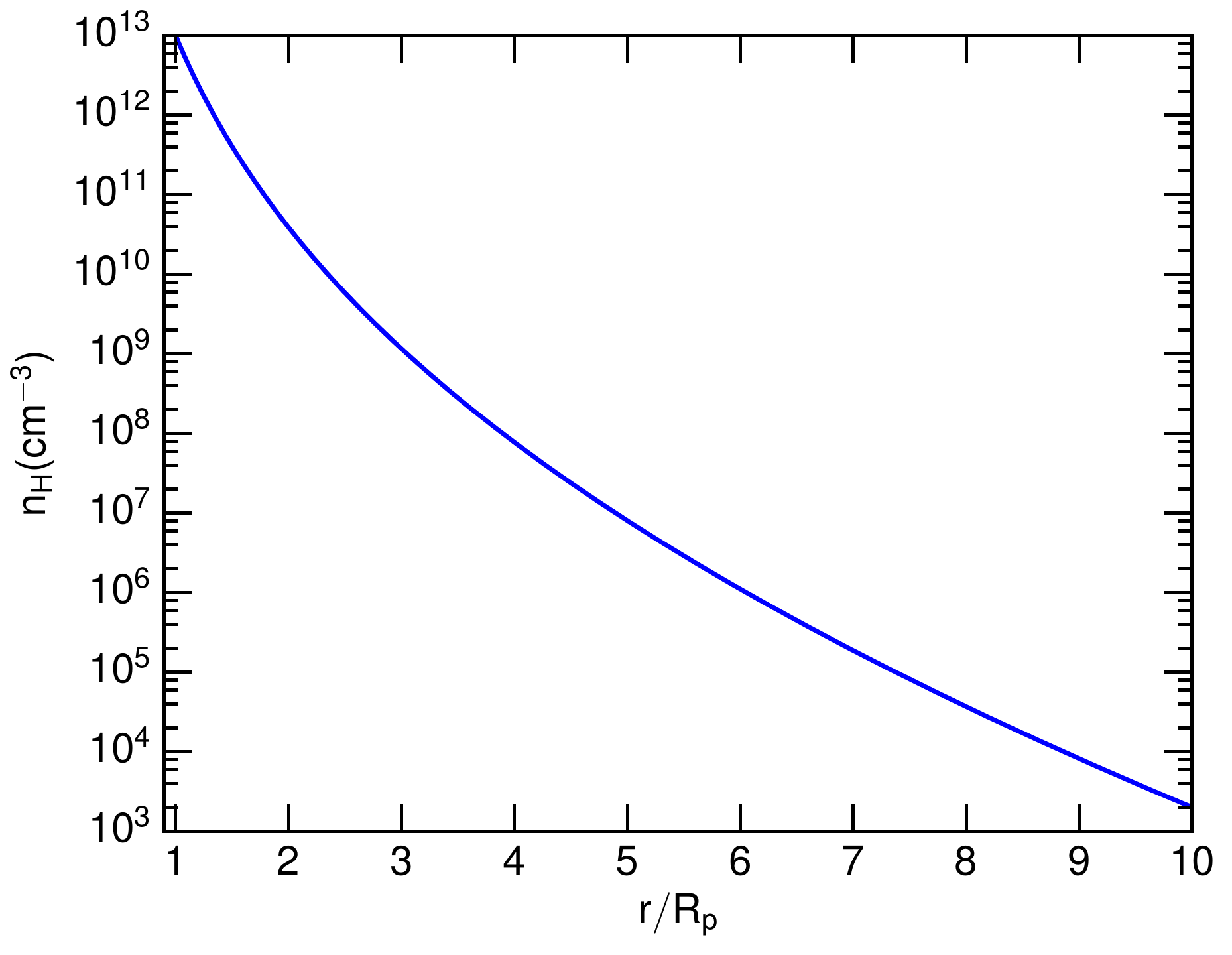}
\caption{The gas number density as a function of the atmosphere altitude $r/R_{\rm p}$. The density at $r=R_{\rm p}$ is chosen to be $n_{\rm H}=n_{\rm p}=10^{13}\cm^{-3}$.}
\label{fig:den}
\end{figure}

The temperature at low altitude near the planet surface can be taken to be the same as the planet surface temperature (see Eq. \ref{eq:Tsf}). At high altitude, the effect of UV radiation is important for gas heating (\citealt{2009ApJ...693...23M}) as well as photoelectric heating of dust by FUV photons \citep{Mitani:2020wa}, and the gas temperature would be much larger. Thus, we assume the upper atmosphere of the hot Jupiter consists of a cool layer of $r<1.1R_{\rm p}$ with $T_{\rm gas}\sim T_{p}\sim 2000\K$ and a warm layer of $T_{\rm gas}\sim 10^{4}\K$. However, the gas temperature only varies by a factor of a few in the atmosphere. Therefore, we adopt the profile of the gas temperature as
\bea
T_{\gas}=T_{\rm p}\left(\frac{r}{R_{\rm p}}\right)^{-\alpha_{T}}
\ena
where $T_{\rm p}$ is the gas temperature at the planet surface of $r=1.1R_{\rm p}$ For hot Jupiters, we assume $T_{\rm p}\approx 10^{4}\K$ and $\alpha_{T}\approx -0.7$ (\cite{2009ApJ...693...23M}) or \cite{Mitani:2020wa}).

Table \ref{tab:model} presents the model adopted for our calculations in this paper, including the stellar effective temperature, stellar radius, planet radius and mass in units of the Jupiter radius and mass ($R_{J}, M_{J}$), the magnetic field strength ($B$).

\begin{table}
\caption{Hot Jupiter Model Parameters}\label{tab:model}
\begin{tabular}{l l} \hline\hline
{Parameters} & Values\cr
\hline\\
Stellar radius & $R_{\star}=1R_{\odot}$ \cr
Stellar mass & $M_{\star}=1M_{\odot}$\cr
Stellar temperature& $T_{\star}=7000\K$ \cr
Planet radius & $1.4R_{J}$\cr
Planet mass & $M_{J}$  \cr
Planet orbital distance & $D_{\rm p}=0.05, 0.1, 0.5, 1\AU$\cr
Number density at $r=R_{\rm p}$ & $n_{\rm p}=10^{13}\cm^{-3}$\cr
Gas temperature at $r=1.1R_{\rm p}$ & $T_{\rm p}=10^{4}\K$\cr
Magnetic field strength  & $B=10$ G\cr

\hline
\hline
\end{tabular}
\end{table}

Figure \ref{fig:den} shows the variation of the gas density with the altitude, assuming the typical density at $r=R_{\rm p}$ to be $n_{\rm p}=10^{13}\cm^{-3}$, $T_{\rm p}=10^{4}\K$ with $\alpha_{T}=0.5$.

\begin{figure}
\includegraphics[width=0.45\textwidth]{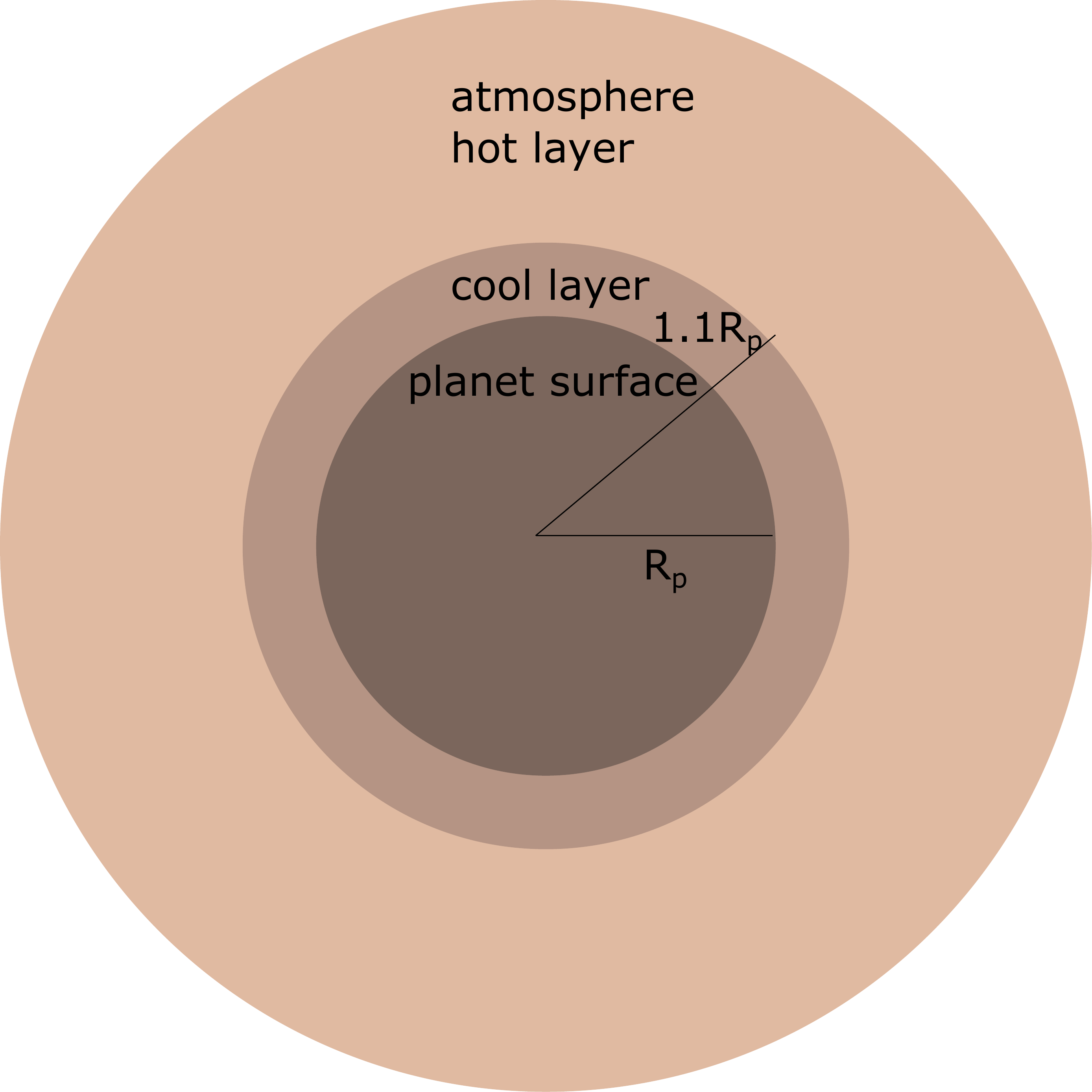}
\caption{Schematic illustration of the hot Jupiter atmosphere. The cool layer is thin of thickness $0.1R_{\rm p}$ and the hot layer extends beyond $1.1R_{\rm p}$.}
\label{fig:schematic}
\end{figure}

\section{Radiative torques of irregular grains}\label{sec:RATs}
In this section, we describe RATs of irregular grains and derive analytical formulae for the RAT efficiency averaged over an arbitrary radiation spectrum, which will be used for alignment and disruption studies.

\subsection{Radiation field}
Let $u_{\lambda}$ be the spectral energy density and $\gamma$ be the anisotropy degree of the radiation field. The anisotropy degree is minimum of $\gamma=0$ for isotropic radiation field and is maximum of $\gamma=1$ for the unidirectional radiation field from a star.

The energy density of the radiation field is then 
\bea
u_{\rad}=\int_{0}^{\infty} u_{\lambda}d\lambda.\label{eq:urad}
\ena

To describe the strength of a radiation field, we introduce the dimensionless parameter $U=u_{\rm rad}/u_{\rm ISRF}$ with $u_{\rm ISRF}=8.64\times 10^{-13}\erg\cm^{-3}$ being the energy density of the average interstellar radiation field (ISRF) in the solar neighborhood as given by \cite{1983A&A...128..212M} (hereafter MMP83). 

The radiation field of the star is given by the luminosity $L=4\pi R_{\star}^{2}\sigma T_{\star}^{4}$. At the typical distance of $D_{\rm p}$, the radiation strength is 
\bea
U=\frac{u_{\rm rad}}{u_{\rm ISRF}}=\frac{L_{\star}}{4\pi d^{2}c u_{\rm ISRF}}=5.2\times 10^{9}\left(\frac{L_{\star}}{L_{\odot}}\right)\left(\frac{D_{\rm p}}{0.1\AU}\right)^{2}.~~
\ena


\subsection{Radiative torques}

Let $a$ be the effective size of the grain which is defined as the radius of the sphere with the same volume as the irregular grain. Radiative torque (RAT) induced by the interaction of an anisotropic radiation field with the irregular grain is defined as
\bea
{\Gamma}_{\lambda}=\pi a^{2}
\gamma u_{\lambda} \left(\frac{\lambda}{2\pi}\right){Q}_{\Gamma},\label{eq:GammaRAT}
\ena
where ${Q}_{\Gamma}$ is the RAT efficiency (\citealt{1996ApJ...470..551D}; \citealt{2007MNRAS.378..910L}).

The magnitude of RAT efficiency can be approximated by a power-law
\bea
Q_{\Gamma}= \alpha\left(\frac{{\lambda}}{a}\right)^{-\eta}\label{eq:QAMO}
\ena
for $\lambda/a\gtrsim 0.1$, where $\alpha$ and $\eta$ are the constants that depend on the grain size, shape, and optical constants. Numerical calculations of RATs for several shapes of different optical constants in \cite{2007MNRAS.378..910L} find the slight difference in RATs among the realization. They adopted the coefficients $\alpha=0.4,\eta=0$ for $a_{\rm trans}<a<\lambda/0.1$, and $\alpha=2.33,\eta=3$ for $a<a_{\rm trans}$ where $a_{\rm trans}=\lambda/1.8$ denotes the transition size at which the RAT efficiency slope changes. Thus, the maximum RAT efficiency is $Q_{\Gamma,\max}=\alpha$. 

An extensive study for a large number of irregular shapes by \cite{2019ApJ...878...96H} shows moderate difference in RATs for silicate, carbonaceous, and iron compositions. Moreover, the analytical formula (Equation \ref{eq:QAMO}) is also in good agreement with their numerical calculations. Therefore, one can use Equation (\ref{eq:QAMO}) for the different grain compositions and grain shapes, and the difference is of order unity.

The radiative torque averaged over the incident radiation spectrum of spectral energy density $u_{\lambda}$ is defined as
\bea
\overline{\Gamma}_{\rm RAT}&=&\pi a^{2}
\gamma u_{\rad} \left(\frac{\overline{\lambda}}{2\pi}\right)\overline{Q}_{\Gamma},\label{eq:GammaRAT_num}
\ena
where
\bea
\overline{Q}_{\Gamma} = \frac{\int_{0}^{\infty} \lambda Q_{\Gamma}u_{\lambda} d\lambda}{\int_{0}^{\infty} \lambda u_{\lambda} d\lambda},~\bar{\lambda}=\frac{\int_{0}^{\infty} \lambda u_{\lambda}d\lambda}{u_{\rm rad}}\label{eq:Qavg_num}
\ena
are the average RAT efficiency and the mean wavelength, respectively.

In general, one can numerically calculate the average RAT and average RAT efficiency using Equations (\ref{eq:GammaRAT_num}) and (\ref{eq:Qavg_num}) for an arbitrary radiation spectrum $u_{\lambda}$. To facilitate analysis of grain alignment from observations, in the following, we will derive analytical formulae for $\overline{Q}_{\Gamma}$ and $\bar{\lambda}$ for two popular radiation fields.

\subsection{Average RAT over a stellar radiation spectrum}

For a radiation field produced by a star of temperature $T_{\star}$, the spectral energy density at distance $d$ from the star is
\bea
u_{\lambda}(T_{\star})=\frac{4\pi R_{\star}^{2}F_{\lambda}}{4\pi d^{2}c}=\frac{\pi B_{\lambda}(T_{\star})}{c}\left(\frac{R_{\star}}{d}\right)^{2},
\ena
where $F_{\lambda}=\pi B_{\lambda}$ is the radiation flux from the stellar surface of radius $R_{\star}$, and $B_{\lambda}=(2hc^{2}/\lambda^{5}) (\exp(hc/\lambda kT_{\star})-1)^{-1}$ is the Planck function. For simplicity, we first disregard the reddening effect by intervening dust.

The total radiation energy density of the stellar radiation field becomes
\bea
u_{\rm rad}(T_{\star})=\left(\frac{R_{\star}}{d}\right)^{2}\frac{\int_{0}^{\infty} \pi B_{\lambda}(T_{\star})d\lambda}{c}=\left(\frac{R_{\star}}{d}\right)^{2}\frac{\sigma T_{\star}^{4}}{c},
\ena
where $\sigma=2\pi^{5}k^{4}/(15h^{3}c^{2})$ is the Stefan-Boltzmann constant.

The mean wavelength of the stellar radiation field is given by
\bea
\bar{\lambda}(T_{\star})&&=\frac{\int_{0}^{\infty} \lambda B_{\lambda}(T_{\star})d\lambda}{\int _{0}^{\infty}B_{\lambda}(T_{\star})d\lambda}=
\left(\frac{2\pi k^{3}\Gamma(3)\zeta(3)}{\sigma ch^{2}}\right)\frac{1}{T_{\star}}\nonumber\\
&&\simeq \frac{0.53\cm \K}{T_{\star}},\label{eq:wavemean_star}
\ena
where $\Gamma$ and $\zeta$ are the Gamma and Riemann functions, and we have used the integral formula $\int_{0}^{\infty} x^{s-1}dx/(e^{x}-1)=\Gamma(s)\zeta(s)$ for $s>1$.
 
For small grains of $a<\bar{\lambda}/1.8$, plugging $Q_{\Gamma}$ from Equation (\ref{eq:QAMO}) and $u_{\lambda}(T_{\star})$ into Equation (\ref{eq:Qavg_num}), one obtains the following after taking the integral,
\bea
\overline{Q}_{\Gamma}&=&\frac{2\pi \alpha k^{\eta+3}}{\sigma h^{\eta+2}c^{\eta+1}}\left(\frac{\zeta(3)\Gamma(3)2\pi k^{3}}{\sigma ch^{2}} \right)^{\eta-1}\Gamma(\eta+3)\zeta(\eta+3)\nonumber\\
&&\times\left(\frac{\bar{\lambda}}{a}\right)^{-\eta}.\label{eq:Qmean_star}
\ena

Plugging the RAT parameters of $\alpha=2.33$ and $\eta=3$ into Equation (\ref{eq:Qmean_star}), one obtains
\bea
\overline{Q}_{\Gamma}\simeq 
6\left(\frac{\bar{\lambda}}{a}\right)^{-3}.\label{eq:Qavg_star}
\ena 
Because the average RAT efficiency cannot exceed its maximum RAT efficiency, $Q_{\Gamma,\max}$, the above equation is only valid for grains of size $a\lesssim (Q_{\Gamma,\max}/6)^{1/3}\bar{\lambda}= \bar{\lambda}/2.5$. Large grains of $a>\bar{\lambda}/2.5$ then have $\bar{Q}_{\Gamma}=Q_{\Gamma,\max}=0.4$. Let $a_{\rm trans,\star}\equiv\bar{\lambda}/2.5$ be the transition size of the RAT averaged over the stellar radiation spectrum.


\section{Grain Alignment by Radiative Torques}
\label{sec:RATA}
\subsection{Alignment axis: k-RAT, B-RAT, and E-RAT}
We first describe current understanding of grain alignment by RATs.

An anisotropic radiation field can align dust grains
via the RAT mechanism (see \citealt{2007JQSRT.106..225L}; \citealt{2015ARA&A..53..501A} for reviews). For an ensemble of grains subject to only RATs, a fraction of grains is first spun-up to suprathermal rotation and then driven to be aligned with the ambient magnetic fields (so-called B-RAT) or with the anisotropic radiation direction (i.e., k-RAT) at an attractor point with high angular momentum, usually referred to as high-J attractors. The high-J attractor corresponds to the maximum angular velocity induced by RATs. A large fraction of grains is driven to low-J attractors (\citealt{2007MNRAS.378..910L}; \citealt{2009ApJ...695.1457H}). In the presence of gas collisions, numerical simulations of grain alignment by RATs in \cite{{Hoang:2008gb},{2016ApJ...831..159H}} show that grains at the low-J attractor will be gradually transported to the high-J attractor after several gas damping times, establishing the stable efficient alignment. 

The existence of high-J attractors by RATs is expected for some grain shapes (\citealt{2007MNRAS.378..910L}), and it becomes universal for grains with enhanced magnetic susceptibility via iron inclusions (\citealt{2008ApJ...676L..25L}; \citealt{2016ApJ...831..159H}). This important finding is supported by simulations in \cite{2016ApJ...831..159H} for grains with various magnetic susceptibilities. Recent calculations of RATs for Gaussian random shapes by \cite{Herranen.2021} show a higher fraction of grain shapes with high-J attractors than previously expected by \cite{2019ApJ...878...96H} for ordinary paramagnetic dust. 

\begin{figure}
\includegraphics[width=0.5\textwidth]{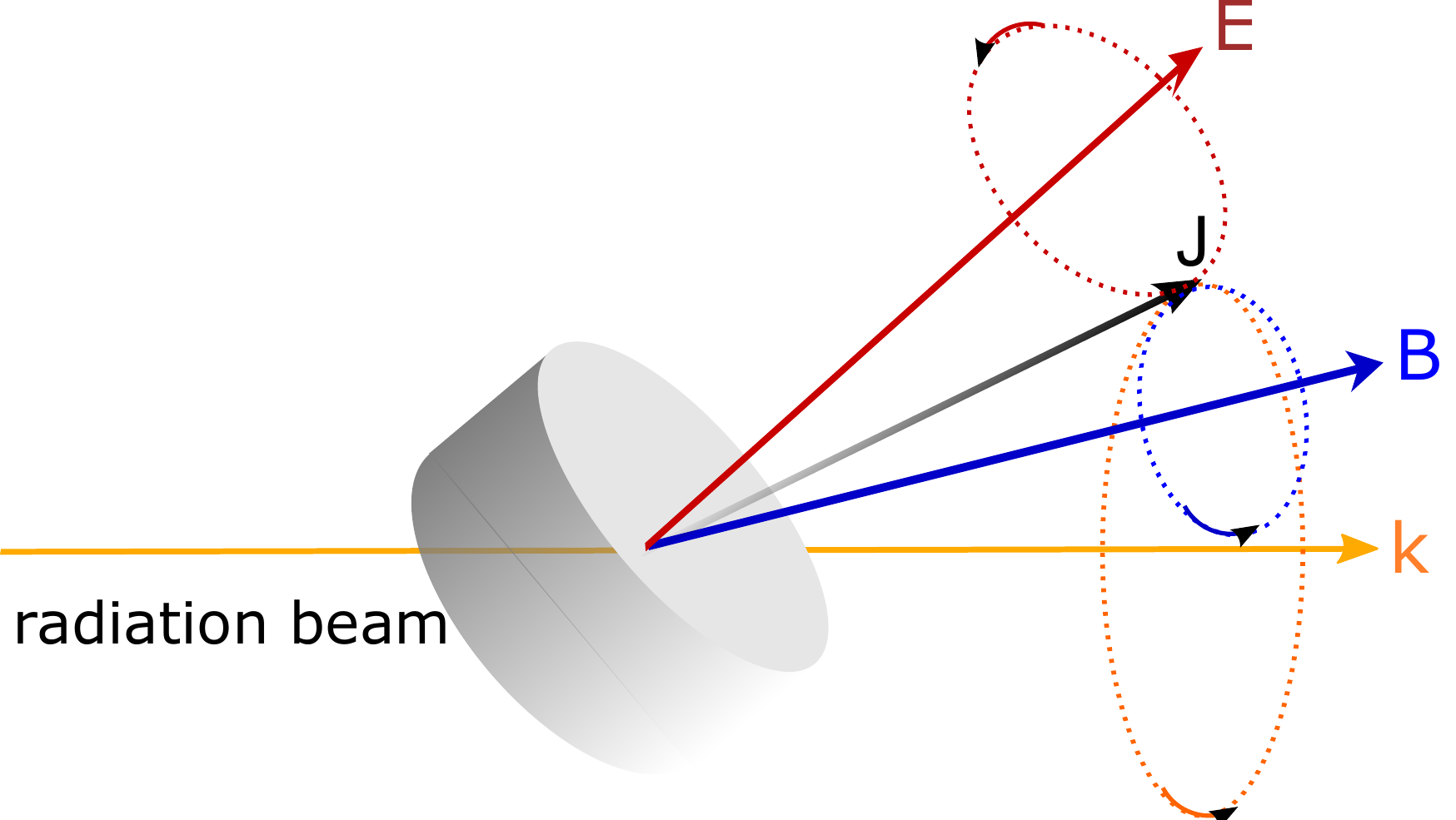}
\caption{Schematic illustration of the precession of the grain angular momentum (${\bf J}$) around three axes, including the radiation direction (${\bf k}$), the magnetic field (${\bf B}$), and the electric field (${\bf E}$). The grain magnetic moment and electric dipole moment are assumed to be along the angular momentum, ${\bf J}$. The fastest precession establishes the alignment axis by RATs.}
\label{fig:KBE_RAT}
\end{figure}

Figure \ref{fig:KBE_RAT} presents the illustration of the grain precession around three axes. The fastest precession establishes the alignment axis by RATs. Thus, grains can be aligned by RATs along the electric field (E-RAT), magnetic field (B-RAT), or radiation direction (k-RAT), depending on their precession timescales. Below, we compare the timescales of these precession processes and determine the alignment axis for the atmosphere conditions.

\subsubsection{Larmor precession}
Silicate grains or grains with iron inclusions are paramagnetic material due to the existence of unpaired electrons. A rotating paramagnetic grain can thus acquire a magnetic moment thanks to the Barnett effect (see \citealt{Barnett:1915p6353}; \citealt{1969mech.book.....L}) and the rotation of its charged body (\citealt{1972MNRAS.158...63M}; \citealt{1976Ap&SS..43..257D}). The Barnett effect, which is shown to be much stronger than the latter, induces a magnetic moment proportional to the grain angular velocity:
\bea
\bmu_{\Bar}=\frac{\chi(0)V\hbar}{g\mu_{B}}\bomega,\label{eq:muBar}
\ena
where $\chi(0)$ is the grain magnetic susceptibility at rest, $g$ is the gyromagnetic ratio, which is $\approx 2$ for electrons, and $\mu_{B}=e\hbar/2m_{e}c$ is the Bohr magneton (see \citealt{Draine:1996p6977} and references therein).

The interaction of this magnetic moment with the external static magnetic field, governed by the torque $[-\bmu_{\Bar}\times \bB]=-\mu_{\Bar}B\sin\xi\hat{\phi}$, causes the regular precession of the grain angular momentum around the magnetic field direction. The rate of such a Larmor precession denoted by $\tau_{B}$, is given by
\bea
\tau_{B}&=&\frac{2\pi}{d\phi/dt}=\frac{2\pi I_{a}\omega}{\mu_{\Bar}B},\nonumber\\
&=&8.4\times 10^{-7}\hat{\rho}^{-1/2}\hat{\chi}^{-1}\hat{B}^{-1} a_{-5}^{2}~\yr,
\label{eq:tauB}
\ena
where $I_{a}=8\pi \rho a^{5}/15$ is the grain inertia moment, $a_{-5}=a/10^{-5}\cm$, and $\hat{B}=B/5$G and $\hat{\chi}=\chi(0)/10^{-4}$ are the normalized magnetic field and magnetic susceptibility, respectively.

Carbonaceous grains are diagmanetic material and have a much smaller magnetic moment due to the attachment of H atoms to the grain surface through hydrogenation (see \citealt{2016ApJ...831..159H}). Since a H electron is already used to make a covalent bond with a C atom, the magnetization is only produced by the H nucleus. The Larmor precession timescale of carbonaceous grains is estimated to be
\bea
\tau_{B,\rm carb}\simeq 1.45\times 10^{-2}\frac{a_{-5}^{2}}{\hat{f}_{p}n_{23}\hat{T}_{d}\hat{B}}~\yr,
\label{eq:tauB_carb}
\ena
where  $n_{23}=n/10^{23}\cm^{-3}$ is the atomic density of material, $\hat{f}_{p}=f_{p}/0.1$ with $f_{p}$ being the fraction of H atoms in this case, $\hat{T}_{d}=T_{d}/20\K$ \citep{2016ApJ...831..159H}, which is much larger than the Larmor precession time of silicate grains (see Eq. \ref{eq:tauB}).

\subsubsection{Grain precession around the anisotropic direction $\kv$}
\cite{2007MNRAS.378..910L} found that the third component of RAT efficiency, $Q_{e3}$, induces the grain precession around the anisotropic direction of radiation field $\kv$. The timescale for such a RAT precession is defined by
\bea 
\tau_{k}&=&\frac{2\pi}{|d\phi/dt|},\nonumber\\
&=& 3.2\times10^{-6}
\hat{\rho}^{1/2}\hat{T}^{1/2}\left(\hat{\lambda}U_{9}\right)^{-1}\hat{Q}_{e3}^{-1}a_{-5}^{1/2}
\yr,\label{eq:tauk}
\ena 
where $\hat{T}=T_{\gas}/100\K$, $U_{9}=U/10^{9}$, and
\bea 
\frac{d\phi}{dt}=\frac{\gamma u_{\rad}\lambda a^{2}}{2I_{a}\omega\sin\xi}{\bf
Q}_{\Gamma}.\hat{\Phi}
=\frac{\gamma u_{\rad}\lambda a ^{2}}{I_{a}\omega} Q_{e3}.\label{eq79}
\ena 
Above, $\hat{Q}_{e3}=Q_{e3}/10^{-2}$ and $\hat{\lambda}=\lambda/1.2 \mum$ for the typical solar radiation field, and the grain angular velocity is taken to be at the thermal velocity, i.e., $\omega=\omega_{T}=(kT_{\rm gas}/I_{a})^{1/2}$. 

\subsubsection{Precession of grain electric dipole moment $\bp$ around $\bE$}

In addition to magnetic moments, dust grains posses electric dipole moments. The electric dipole moment of a grain consists of the intrinsic moment due to molecules and substructures with polar bonds, and the moment due to the asymmetric distribution of grain charge. It can be written as
\bea
p^{2}= (\epsilon Qa)^{2}+p_{\rm int}^{2},\label{eq:ptot}
\ena
where $Q$ is the grain charge and $\epsilon a$ is the displacement between the grain charge centroid and the center of mass (see \citealt{1998ApJ...508..157D}). The grain charge centroid is present for irregular grains even if they perfectly conducting (\citealt{1975duun.book..155P}).

For ultrasmall grains (e.g., polycyclic aromatic hydrocarbons), the intrinsic dipole moment $p_{\rm int}$ dominates, but for larger grains the dipole moment due to the charge distribution becomes dominant. The latter can be rewritten as
\bea
p= 1.0\times 10^{-15} \left(\frac{U}{0.3\rm V}\right)\left(\frac{\epsilon}{10^{-2}}\right)a_{-5}~ {\rm statC} \cm,\label{eq:pE}
\ena
where $U\approx Q/a$ is the grain electrostatic potential.

In an ambient electric field $\bE$, the grain electric dipole moment precesses around $\bE$ at a rate
\bea
\Omega_{E}\equiv\frac{2\pi}{\tau_{E}}=\frac{\bp.\bE}{J},\label{eq:omegaE}
\ena
where $\tau_{E}=2\pi/|d\phi/dt|$ with $\phi$ being the precession angle of $\bJ$ around $\bE$. Here, we have assumed that the dipole moment ${\bf p}$ is coupled to the grain angular momentum, or there is a component of electric field along $\bf J$.

Following \cite{2007JQSRT.106..225L} and \cite{2014MNRAS.438..680H}, the electric field can be produced by static charges in the atmosphere as well as the relative motion of charged grains through the ambient magnetic fields that can occur due to grain radiative acceleration, shocks, gyromotion of the grain in the turbulent magnetic field (see also \citealt{Lazarian.2020bwy}; \citealt{Lazarian.2020} for detailed discussions
). Hot Jupiter exoplanets experience mass loss due to gas-driven or dust-driven radiation pressure winds, with the expansion velocity of the gas of $v_{\rm exp}\sim 10$ km/s (\citealt{2009ApJ...693...23M}). Assuming the coupling of dust and gas, then, dust grains drift through the magnetic field with $v_{\rm grain}=v_{\rm exp}$ induces an electric field of
\bea
E_{\rm ind}&=&\frac{1}{c}v_{\rm grain}B\nonumber\\
&\simeq &3.3\times 10^{-4}\left(\frac{v_{\rm grain}}{10~\rm km\s^{-1}}\right)\left(\frac{B}{10 G}\right) ~{\rm stat~V}\cm^{-1}.
\label{eq:Eind}
\ena

Assuming the electric field $E =0.1 \V\cm^{-1}=0.1\times 10^{8}/c~ {\rm statV cm^{-1}}$ and the grain electrostatic potential $U=0.3 \V$ for Equation (\ref{eq:omegaE}), we obtain
\bea
\tau_{E}&=&7.0 \hat{\rho}\left(\frac{\epsilon}{10^{-2}}\right)^{-1}
\left(\frac{U}{0.3 \rm V}\right)^{-1}\left(\frac{E}{0.1\rm V\cm^{-1}}\right)^{-1}\nonumber\\
&\times&\left(\frac{\omega}{\omega_{T}}\right)
\left(\frac{T_{\gas}}{100\K}\right)^{1/2}a_{-5}^{1/2}\yr.~~~~ \label{eq:tauE}
\ena

\subsubsection{k-RAT vs. B-RAT vs. E-RAT alignment}
The alignment of grains in atmospheres can occur with the radiation direction (k-RAT), the magnetic field (B-RAT), or electric field (E-RAT), depending on the rate of grain precession around these axes (see Figure \ref{fig:KBE_RAT}). From Equations (\ref{eq:tauk}) and (\ref{eq:tauB}), the ratio of the precession rate around the radiation to that around the magnetic field is equal to
\bea  
\frac{\tau_{k}}{\tau_{B}}&=3.3\hat{\rho}\hat{\chi}\hat{T}^{1/2}a^{-3/2}_{-5}\left(\frac{\hat{B}}
{\hat{\lambda}U_{9}}\right)\hat{Q}_{e3}^{-1},\label{eq:taukB}
\ena
for silicate grains.

The magnetic field strength in exoplanetary atmospheres is uncertain, but one expect large magnetic field in hot Jupiter due to its proximity to the star. Simulations (\citealt{Rogers.2014}) and observational measurements by \cite{Cauley.2019} shows $B\sim 10-20 $G. It can be checked easily that for the atmosphere of $U\sim 10^{9}$, $\tau_{k}/\tau_{B} \sim 30$ for $B\sim 10$ G, radiation precession is longer than the Larmor precession. Therefore, the B-RAT alignment is expected for silicate grains or iron inclusions with high magnetic susceptibility. 

For carbonaceous grains, the Larmor precession time is four orders of magnitude larger (see Eq. \ref{eq:tauB_carb}), therefore, grain alignment occurs via k-RAT instead of B-RAT (see also \citealt{Lazarian.2020bwy}; \citealt{Lazarian.2020}).

Similarly, the alignment of silicate grains with the magnetic field (B-RAT) with electric field (E-RAT) depends on the ratio of their precession timescales:
\bea
\frac{\tau_{B}}{\tau_{E}}&=&1.2\times 10^{-7}\hat{\rho}^{-3/2}\hat{\chi}^{-1}\hat{B}^{-1} a_{-5}^{3/2}\left(\frac{\epsilon}{10^{-2}}\right)
\left(\frac{U}{0.3 \rm V}\right)\nonumber\\
&\times&\left(\frac{E}{0.1\rm V\cm^{-1}}\right)
\left(\frac{\omega}{\omega_{T}}\right)^{-1}\hat{T}^{1/2},~~~~ \label{eq:tauE_tauB}
\ena
for silicate grains, which implies the subdominance of E-precession to Larmor precession, and silicate grains experience B-RAT instead of E-RAT.

\subsection{Maximum rotation speed induced by RATs}
For a radiation source with constant luminosity considered in this paper, radiative torques $\Gamma_{\rm RAT}$ are constant and so that the grain angular velocity steadily increases over time. The equilibrium rotation can be achieved at (see \citealt{2007MNRAS.378..910L}; \citealt{2009ApJ...695.1457H}; \citealt{2014MNRAS.438..680H}):
\bea
\omega_{\rm RAT}=\frac{\Gamma_{\rm RAT}\tau_{\rm damp}}{I_{a}},~~~~~\label{eq:omega_RAT0}
\ena
where $I_{a}=8\pi \rho a^{5}/15$ is the principal moment of inertia of spherical grains of radius $a$, and $\tau_{\rm damp}$ is the rotational damping time.

The total rotational damping rate by gas collisions and IR emission can be written as
\bea
\tau_{\rm damp}^{-1}=\tau_{\gas}^{-1}(1+ F_{\rm IR}),\label{eq:taudamp}
\ena
where $\tau_{\rm gas}$ is the damping time due to gas collisions followed by evaporation and given by
\bea
\tau_{\gas}&=&\frac{3}{4\sqrt{\pi}}\frac{I_{a}}{1.2n_{\rm H}m_{\rm H}
v_{\rm th}a^{4}}\nonumber\\
&\simeq& 0.26a_{-5}\hat{\rho}\left(\frac{1}{n_{10}T_{3}^{1/2}}\right)~{\rm yr},~~
\label{eq:taugas}
\ena
where $a_{-5}=a/(10^{-5}\cm)$, $\hat{\rho}=\rho/(3\g\cm^{-3})$ with $\rho$ being the dust mass density, $v_{\rm th}=\left(2k_{\B}T_{\rm gas}/m_{\rm H}\right)^{1/2}$ is the thermal velocity of gas atoms of mass $m_{\rm H}$, and the spherical grains are assumed (\citealt{2009ApJ...695.1457H}; \citealt{1996ApJ...470..551D}). The gas temperature and gas density are normalized to their typical values of the atmosphere, such that $T_{3}=T_{\gas}/10^{3}\K$, and $n_{10}=n_{\H}/(10^{10}\cm^{-3})$
respectively.

The second term in Equation (\ref{eq:taudamp}), $\tau_{\rm IR}^{-1}=F_{\rm IR}\tau_{\rm gas}^{-1}$, is the rotational damping due to 
IR photons emitted by the grain carry away part of the grain's angular. For strong radiation fields or not very small sizes, grains can achieve equilibrium temperature, such that the IR damping coefficient (see \citealt{1998ApJ...508..157D}) can be calculated as
\bea
F_{\rm IR}\simeq \left(\frac{3.8\times 10^{-4}}{a_{-5}}\right)\left(\frac{U_{9}^{2/3}}{n_{10}T_{3}^{1/2}}\right).\label{eq:FIR}
\ena 
For high density of $U_{9}/n_{10}T_{3}<1$, the IR damping is subdominant, but it becomes important when the density decreases at higher altitude.

Plugging $\tau_{\rm damp}$ and $\overline{\Gamma}_{\rm RAT}$ with $Q_{\rm RAT}$ from Equation (\ref{eq:Qavg_star}) into Equation (\ref{eq:omega_RAT0}), one obtains
\bea
\omega_{\rm RAT}&=& \frac{3\gamma u_{\rm rad}a\bar{\lambda}^{-2}}{1.6n_{\rm H}\sqrt{2\pi m_{\rm H}kT_{\rm gas}}}\left(\frac{1}{1+F_{\rm IR}}\right)\nonumber\\
&\simeq &9.4\times 10^{6} a_{-5}\left(\frac{\bar{\lambda}}{1.2\mum}\right)^{-2}
\left(\frac{\gamma U}{n_{3}T_{1}^{1/2}}\right)\nonumber\\
&&\times\left(\frac{1}{1+F_{\rm IR}}\right)\rad\s^{-1},\label{eq:omega_RAT1}
\ena
for grains with $a\lesssim a_{\rm trans, \star}$, and
\bea
\omega_{\rm RAT}&=&\frac{1.5\gamma u_{\rm rad}\bar{\lambda}a^{-2}}{16n_{\rm H}\sqrt{2\pi m_{\rm H}kT_{\rm gas}}}\left(\frac{1}{1+F_{\rm IR}}\right)\nonumber\\
&\simeq& 8.1\times 10^{8}a_{-5}^{-2}\left(\frac{\bar{\lambda}}{1.2\mum}\right) \left(\frac{\gamma U}{n_{3}T_{1}^{1/2}}\right)\nonumber\\
&&\times\left(\frac{1}{1+F_{\rm IR}}\right)\rad\s^{-1},\label{eq:omega_RAT2}
\ena
for grains with $a> a_{\rm trans, \star}$. Here $a_{-5}=a/10^{-5}\cm$, $n_{3}=n_{\H}/10^{3}\cm^{-3}$, $T_{1}=T_{\gas}/10\K$, $F_{\rm IR}$ is the IR damping coefficient (see Eq. \ref{eq:FIR}) (e.g., \citealt{1996ApJ...470..551D}). The stellar radiation field has $\gamma=1$.

\subsection{Grain Alignment Size}

Efficient grain alignment is achieved only when grains can rotate with angular velocity greater than its thermal value, which is termed suprathermal rotation. The grain thermal angular velocity is
\bea
\omega_{T}&=&\left(\frac{kT_{\rm gas}}{I_{a}}\right)^{1/2}=\left(\frac{15kT_{\gas}}{8\pi\rho a^{5}}\right)^{1/2}\nonumber\\
&\simeq &5.23\times 10^{5}\hat{\rho}^{-1/2}a_{-5}^{-5/2}T_{3}^{1/2}~ \rm rad\s^{-1}.\label{eq:omega_th}
\ena

Using the suprathermal rotation threshold of $\omega_{\rm RAT}(a_{\rm align}) = 3\omega_{T}$ as in \cite{2008MNRAS.388..117H}, one obtains the minimum size of aligned grains (hereafter alignment size) as follows:
\bea
a_{\rm align}&=&\left(\frac{1.6n_{\rm H}T_{\rm gas}}{\gamma u_{\rm rad}\bar{\lambda}^{-2}} \right)^{2/7}\left(\frac{15m_{\rm H}k^{2}}{4\rho}\right)^{1/7}(1+F_{\rm IR})^{2/7}\nonumber\\
&\simeq &0.029\hat{\rho}^{-1/7} \left(\frac{\gamma U_{9}}{n_{10}T_{3}}\right)^{-2/7} \nonumber\\
    &&\times \left(\frac{\bar{\lambda}}{1.2\mum}\right)^{4/7} (1+F_{\rm IR})^{2/7} ~\mum,\label{eq:aalign_ana}
\ena 
which implies $a_{\rm align} \sim 0.055 \mum$ for a dense cloud of $n_{\H}=10^{3}\cm^{-3}$ exposed to the local radiation field of $\gamma=1$, $U=10^{9}$, and $\bar{\lambda}=1.2\mum$. In general, grain alignment by RATs depends on five local physical parameters of the gas ($n_{\H}, T_{\gas}$) and the radiation field at the cloud surface ($\gamma, U,\bar{\lambda}$). The alignment size increases with with increasing $n_{\rm H}$ and decreasing $U$.

For low altitude atmospheres with high gas density, the gas damping dominates over IR damping, and $F_{\rm IR}\ll 1$. Thus, the alignment size can be analytically obtained from Equation (\ref{eq:aalign_ana}) when $n_{\rm H}, T_{\rm gas}$, and $T_{d}$ are known. At high altitude where $n_{H}$ is smaller, the contribution of IR damping becomes important. Thus, one must solve Equation numerically for the alignment size due to the dependence of $F_{\rm IR}$ on the grain size (see Eq.\ref{eq:FIR}). We calculate $a_{\rm align}$ using both analytical equations with $F_{\rm IR}=0$ and numerical calculations taking into account the arbitrary $F_{\rm IR}$.

\begin{figure}
\includegraphics[width=0.5\textwidth]{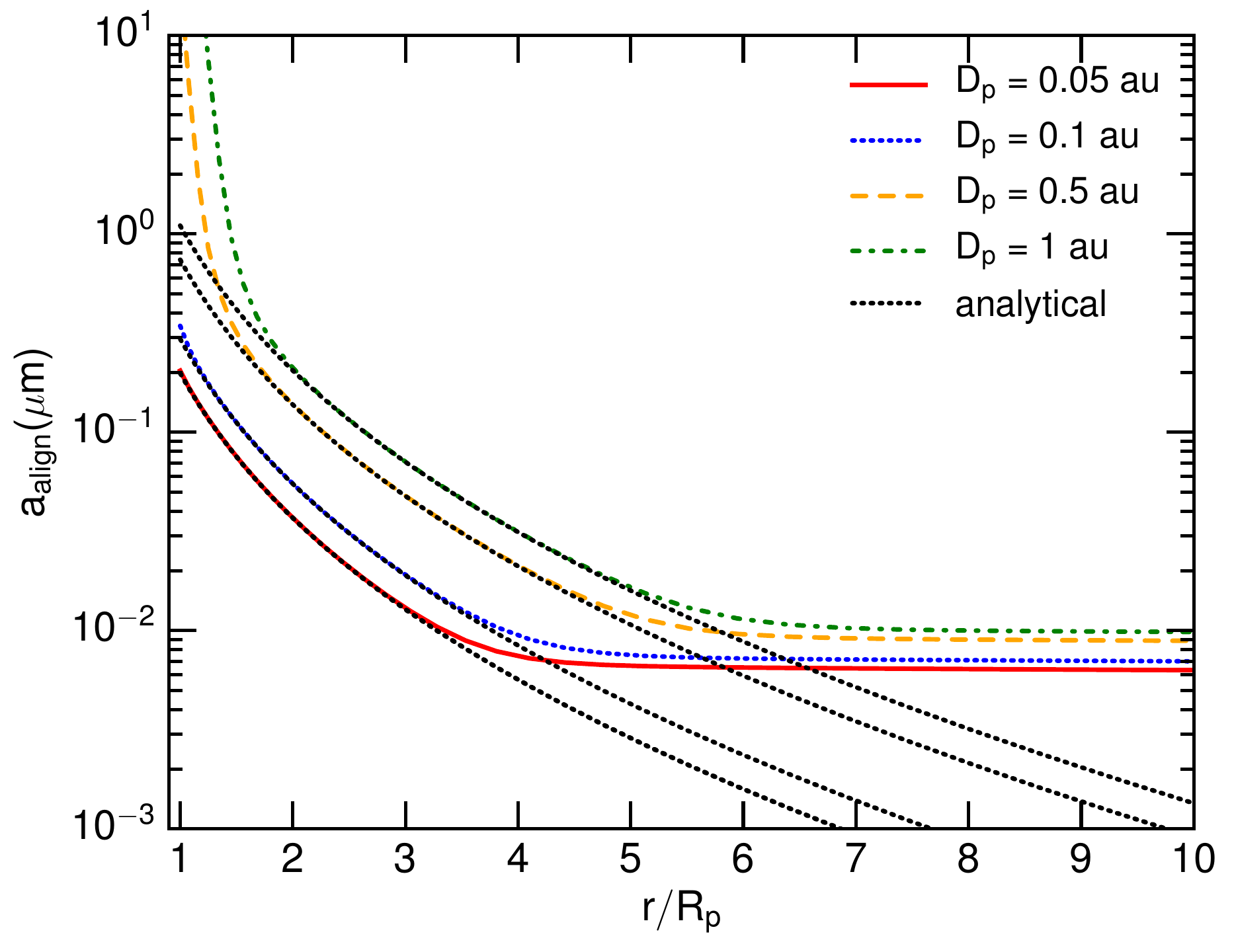}
\caption{Alignment size as a function of the altitude for the different distances. Analytical results (dotted black lines) given by Equations \ref{eq:aalign_ana} with $F_{\rm IR}=0$ are in good agreement with numerical results for low altitudes. Small grains can be aligned by RATs at altitude of $r/R_{\rm p}>2$.}
\label{fig:align}
\end{figure}

Figure \ref{fig:align} shows the minimum size of aligned grains by RATs as a function of the altitude for the different planet distances to the central star. For a given typical distance of the exoplanet at $D_{\rm p}=0.05$ au, the alignment size decreases with increasing the altitude increases due to the rapid decrease of the gas density $n_{\H}$ (see Figure \ref{fig:den}). At high altitude, the alignment size becomes saturated because the rotational damping is now dominated by IR emission. For a given altitude, the alignment size is larger for larger distance $D_{\rm p}$ due to the decrease of the radiation intensity.

\section{Grain Disruption by Radiative Torques}
\label{sec:RATD}
\subsection{The RATD mechanism}
A rotating grain at angular velocity $\omega$ develops a centrifugal stress $S=\rho a^{2}\omega^{2}/4$. When the grain rotation rate is sufficiently high, such that the tensile stress can exceed the tensile strength, $S_{\rm max}$, the grain is instantaneously disrupted into fragments. This mechanism is termed RAdiative Torque Disruption (RATD; \citealt{Hoang:2019da}). The critical angular velocity is obtained by setting $S=S_{\rm max}$:
\bea
\omega_{\rm disr}&=&\frac{2}{a}\left(\frac{S_{\max}}{\rho} \right)^{1/2}\nonumber\\
&\simeq& \frac{3.65\times 10^{8}}{a_{-5}}\hat{\rho}^{-1/2}S_{\max,7}^{1/2}~\rad\s^{-1},\label{eq:omega_cri}
\ena
where $S_{\max,7}=S_{\max}/(10^{7}\erg\cm^{-3})$.

The tensile strength of dust is uncertain, depending on grain structure (compact vs. composite) and composition \citep{2019ApJ...876...13H}. For solid (compact) grains or dense aggregates, one expect $S_{\rm max}> 10^{7}\erg\cm^{-3}$, and a lower tensile strength of $S_{\rm max}<10^{6}\erg\cm^{-3}$ is expected for for the porous group (see \citealt{2019A&A...630A..24G}). In exoplanetary atmospheres, grains can grow to large sizes and have composite structures as a result of coagulation process, resulting in a low tensile strength.

\subsection{Disruption size}

Comparing Equations (\ref{eq:omega_RAT1}) and (\ref{eq:omega_cri}), one can obtain the disruption grain size (see also \citealt{Hoang.2021}):
\bea
a_{\rm disr}&=&\left(\frac{3.2n_{\rm H}\sqrt{2\pi m_{\rm H}kT_{\rm gas}}}{3\gamma u_{\rm rad}\bar{\lambda}^{-2}}\right)^{1/2}\left(\frac{S_{\rm max}}{\rho}\right)^{1/4}(1+F_{\rm IR})^{1/2}\nonumber\\
&\simeq& 0.197 \left(\frac{\gamma U_{9}}{n_{10}T_{3}^{1/2}}\right)^{-1/2}\left(\frac{\bar{\lambda}}{1.2\mum}\right) \hat{\rho}^{-1/4}S_{\max,7}^{1/4}\nonumber\\
&&\times (1+F_{\rm IR})^{1/2}\mum,
\label{eq:adisr_ana}
\ena
which depends on the local gas properties, radiation field, and the grain tensile strength. 

Due to the decrease of the grain rotation rate for $a>a_{\rm trans,\star}$ (see Eq. \ref{eq:omega_RAT2}), the rotational disruption occurs only if $a_{\rm disr}<a_{\rm trans, \star}$. In this case, there exists a maximum size of grains that can still be disrupted by centrifugal stress (\citealt{2020ApJ...891...38H}),
\bea
a_{\rm disr,max}&=&\frac{3\gamma u_{\rm rad}\bar{\lambda}}{64n_{\rm H}\sqrt{2\pi m_{\rm H}kT_{\rm gas}}}\left(\frac{S_{\rm max}}{\rho}\right)^{-1/2}(1+F_{\rm IR})^{-1}\nonumber\\
&\simeq& 
2.2\left(\frac{\gamma U_{9}}{n_{10}T_{3}^{1/2}}\right)\left(\frac{\bar{\lambda}}{1.2\mum}\right)\hat{\rho}^{1/2}S_{\max,7}^{-1/2}\nonumber\\
&\times&(1+F_{\rm IR})^{-1}\mum.\label{eq:adisr_up}
\ena 

Thus, the disruption size can be analytically obtained from Equation (\ref{eq:adisr_ana}) when $n_{\rm H}, T_{\rm gas}$, and $T_{d}$ are known. At high altitude where $n_{\rm H}$ is smaller, the contribution of IR damping becomes important. Thus, one must solve Equation numerically for the disruption size due to the dependence of $F_{\rm IR}$ on the grain size (see Eq.\ref{eq:FIR}). We calculate $a_{\rm disr}$ and $a_{\rm disr,max}$ using both analytical equations with $F_{\rm IR}=0$ and numerical calculations taking into account the arbitrary $F_{\rm IR}$.

\begin{figure*}
\includegraphics[width=0.5\textwidth]{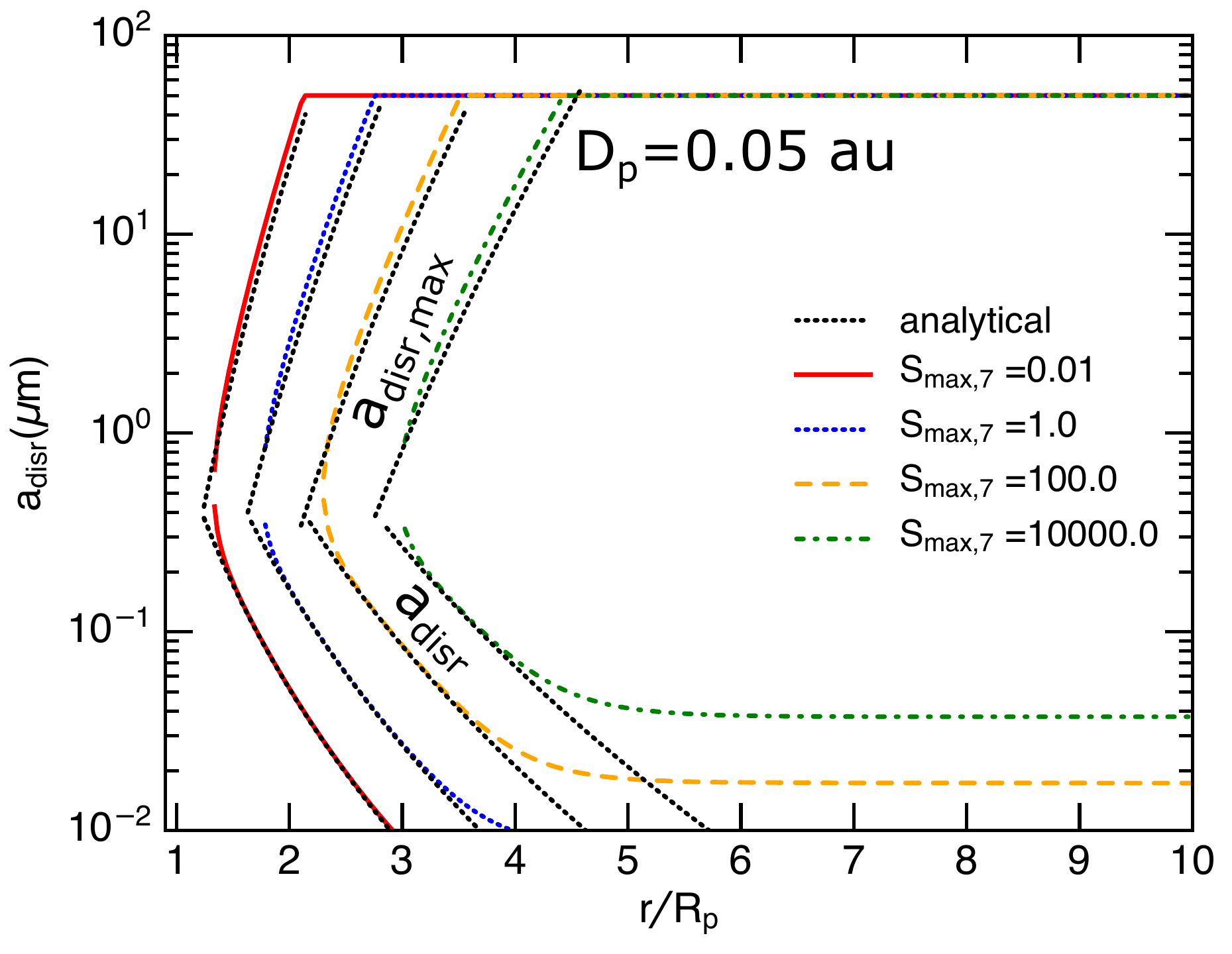}
\includegraphics[width=0.5\textwidth]{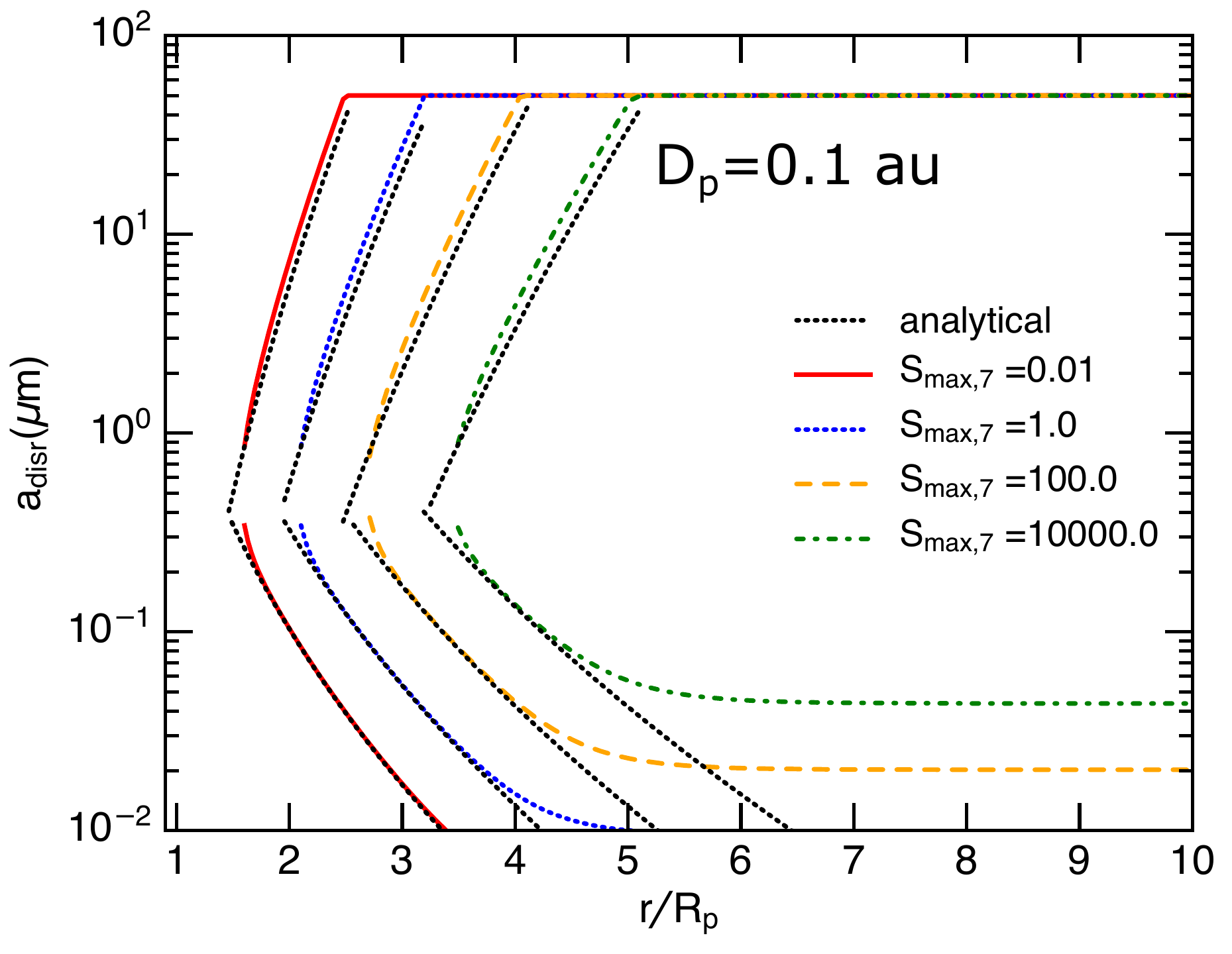}
\includegraphics[width=0.5\textwidth]{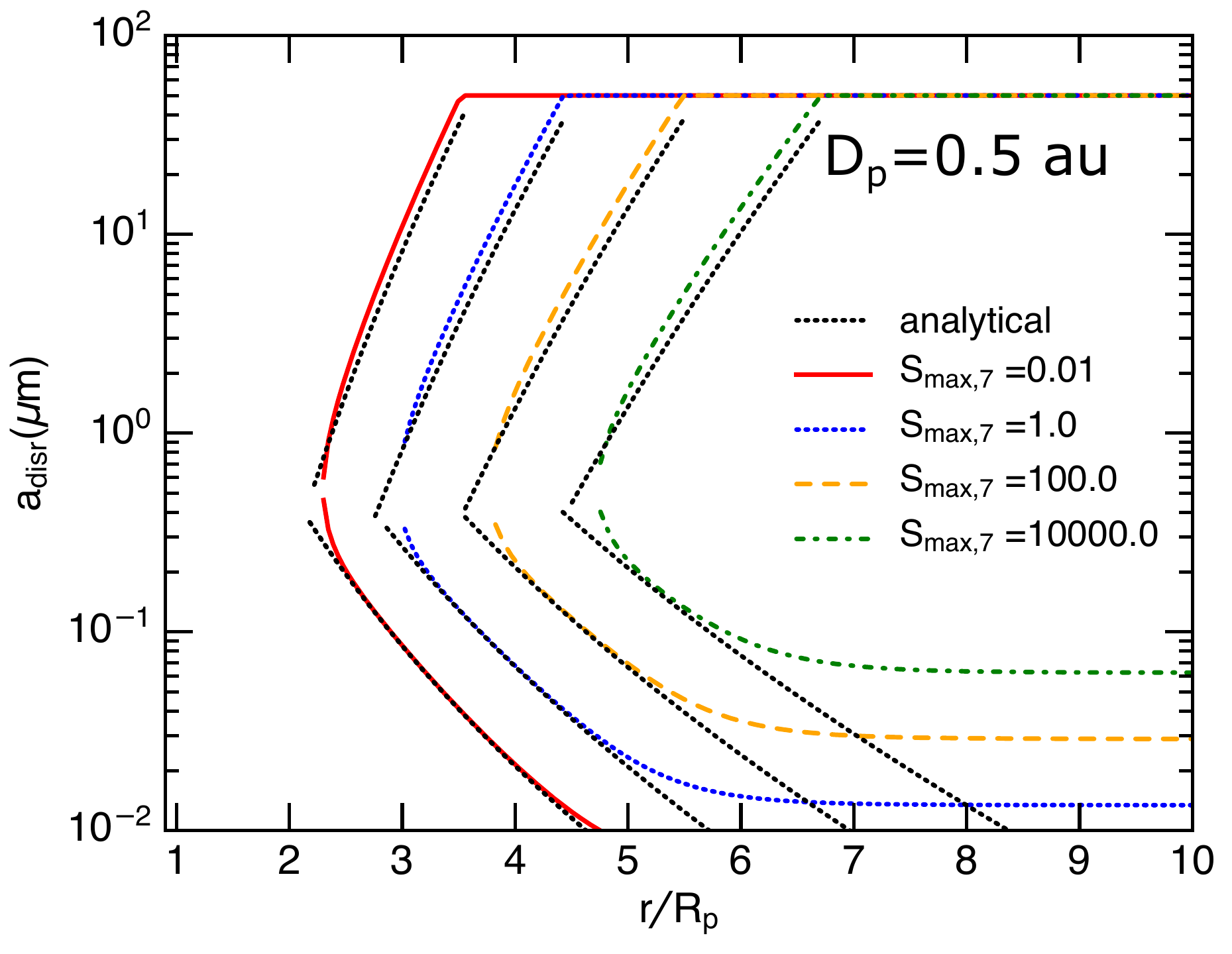}
\includegraphics[width=0.5\textwidth]{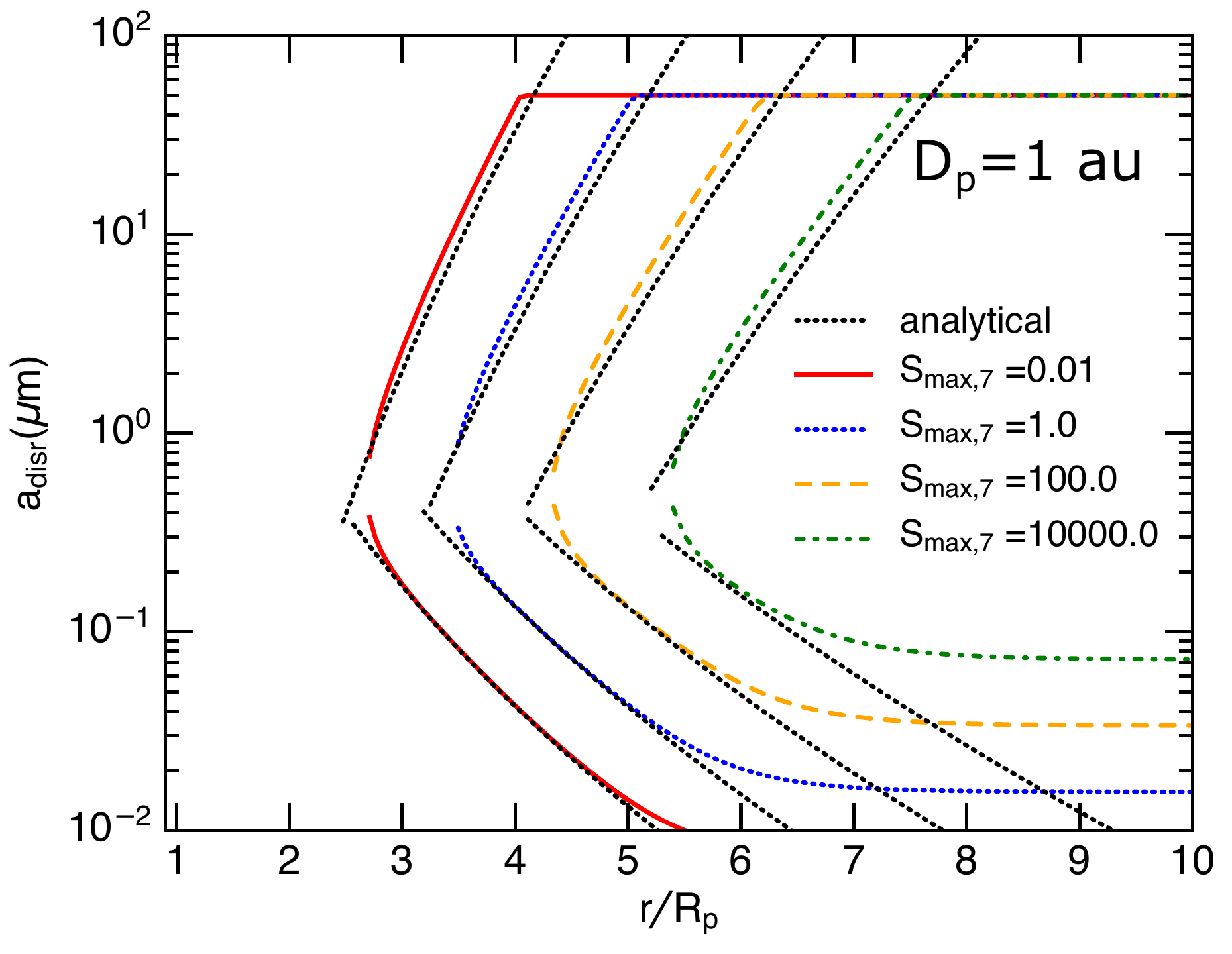}
\caption{The disruption size as a function of the altitude for the different distances from $D_{\rm p}=0.05-1$ au and different tensile strengths of $S_{\max}$. Large grains can be disrupted at altitude of $r/R_{\rm p}>2$. Analytical results (dotted black lines) given by Equations \ref{eq:adisr_ana} with $F_{\rm IR}=0$ are in good agreement with numerical results for low altitudes.}
\label{fig:adisr}
\end{figure*}

Figure \ref{fig:adisr} shows the minimum size and maximum size of grains disrupted by RATs as a function of the altitude for the different tensile strengths. For a given typical distance of the exoplanet at $D_{\rm p}=0.05$ au, the RATD effect is not effective near the planet surface due to extremely high density. Grains have low tensile strength is disrupted stronger by RATD. Even compact grains of highest tensile strength of $S_{\max}=10^{11}\erg\cm^{-3}$ could be still disrupted. As the altitude increases, the RATD effect becomes more efficient due to the decrease of the gas density (see Figure \ref{fig:den}), and it achieves the saturated values when the IR damping becomes dominant over gas collisions. The RATD effect is more efficient for the smaller distance of exoplanets to the host star. For grains with low tensile strength of $S_{\max}=10^{7}\erg\cm^{-3}$, the RATD effect is efficient when the exoplanet is at distance of $D_{\rm p}\sim 1$ au.

The RATD effect occurs only at $r>2R_{p}$ when the gas density drops. Therefore, the disruption occurs efficiently in the upper atmosphere starting from the photoionization base of $1.1R_{p}$ (see \cite{2009ApJ...693...23M}).

\section{Discussion}\label{sec:discuss}

\subsection{Dust polarization and tracing magnetic fields in hot Jupiters}
Dust polarization induced by aligned grains is widely used to study interstellar magnetic fields (\citealt{2012ARA&A..50...29C}; \citealt{2019FrASS...6...15P}) and to constrain dust properties (\citealt{Ngoc.2021}; \citealt{2z1}). 
Following the RAT paradigm, we studied grain alignment by RATs and discussed three possible alignment scenarios in the atmospheres of hot Jupiters. For silicate grains or grains with iron inclusions (\citealt{2016ApJ...831..159H}), they can be aligned with the magnetic field (B-RAT). In contrast, carbonaceous grains can be aligned with the radiation direction (k-RAT) due to their negligible magnetic susceptibility. Due to the expansion of atmospheric gas in hot Jupiters, dust grains can drift across the magnetic field and induce an induced electric field. The grains thus can experience the alignment along the electric field which is perpendicular to the magnetic field, so-called E-RAT alignment. 

The B-RAT alignment induces the polarization of the host star as well as polarized thermal emission. For carbonaceous grains with k-RAT, polarization produced by differential absorption by aligned grains is inefficient because the short axis is along the radiation which is the line of sight. Therefore, observations of dust polarization would be crucial to test k-RAT vs. B-RAT alignment mechanisms.

Moreover, for the B-RAT alignment, the optical-UV polarization vectors are along the magnetic field, and thermal dust polarization vectors are perpendicular to the magnetic field. Therefore, one can infer the magnetic field using dust polarization. The optical-NIR polarization is contaminated with dust scattering when the planet makes an angle with the host star, so the most plausible way is to use polarized thermal dust emission. Due to high dust temperature, polarized thermal emission is peaked at $\lambda_{\rm peak}\sim 20\mum (1000\K/T_{d})$.

Recently, \cite{JensenClem:2020do} and \cite{Holstein.2021} presented a survey of NIR polarized thermal emission from brown dwarfs and its planet companions. However, this NIR polarized signal is due to scattering of starlight on its atmosphere and aligned dust grains, which makes it difficult to disentangle the polarization by these two mechanisms.

Furthermore, circular polarization (CP) produced by scattering of starlight by aligned grains is also a tracer of grain alignment (CP; \citealt{2014MNRAS.438..680H}). Indeed, if grains experience k-RAT alignment, the degree of CP is negligible because the incident radiation direction is the same as the alignment axis, whereas B-RAT and E-RAT alignment induces considerable CP.

\subsection{Grain rotational disruption and the grain size distribution}
Dust formation in hot Jupiter atmospheres is studied in detail. Various grain size distributions are found for dust in atmospheres, with the different upper cutoff of the grains size distribution (see \citealt{Gao.2021}). \cite{Ohno:2020bj}. Using the RATD mechanism, we find that large grains can be disrupted by RATD into smaller sizes at altitude of $r>3R_{\rm p}$, but weak grains can be disrupted at lower altitudes. The presence of small fragments increases the UV-optical extinction and reproduces the flat transmission at $\lambda<1\mum$.

Hot Jupiters are thought to form due to the migration of giant gas planets from outer distances inward. During this process, dust grains are gradually disrupted by RATs and large grains are depleted and small grains are more abundant. Therefore, the grain size distribution of atmosphere dust will be shifted to smaller sizes during the migration process. This prediction would be tested with observations.

\subsection{Implications for high-altitude clouds in hot Jupiters and super-puffs}
The presence of high-altitude clouds implied by the flat transmission spectrum is a puzzle because dust cannot form at high altitudes. Photochemical haze can reproduce the flat spectrum if the haze formation is sufficiently fast, but its efficiency is still uncertain (\citealt{Lavvas.2017}; \citealt{He:2020eh}). 
Recently, \cite{Ohno:2020bj} suggested that the formation of dust aggregates with low volume density can easily escape from the gravity to float to high altitude. However, their dust aggregate implies the spectral slope of $\alpha=-2$, which is much larger than the observed slope of $\alpha=-5$. Therefore, the compact-dust model is not ruled out. 

Several models are proposed to explain the super-puffs, including dusty outflows (\citealt{j82}), dust formation by photochemical haze (\citep{Gao:2020dq}; \citealt{Ohno.2021}). Small dust grains are thought to cause the atmosphere escape and winds in super-puffs \cite{Ohno.2021}. 

Following \cite{2019ApJ...874..159T}, the tensile strength of dust aggregates decreases with increasing the monomer radius and can be fitted with an analytical formula
\bea
S_{\max} &\simeq& 9.51\times 10^{4} \left(\frac{\gamma_{\rm sf}}{100\erg\cm^{-2}}\right) \nonumber\\
&\times&\left(\frac{r_{0}}{0.1\mum}\right)^{-1}\left(\frac{\phi}{0.1}\right)^{1.8} \erg\cm^{-3},\label{eq:Smax}
\ena
where $\gamma_{\rm sf}$ is the surface energy per unit area of the material, $r_{0}$ is the monomer radius, and $\phi$ is the volume filling factor of monomers (see also \citealt{2020MNRAS.496.1667K}). For large composite grains made of monomers of radius $r_{0}=0.1\mum$  with $\phi=0.1$ and $\gamma_{\rm sf}=100\erg\cm^{-2}$, Equation (\ref{eq:Smax}) implies $S_{\max}\approx 10^{5}\erg\cm^{-3}$. 

With the low tensile strength of $S_{\max}$ implied by Equation (\ref{eq:Smax}), our calculations show that dust aggregates are efficiently disrupted by RATD into small compact fragments (see Figure \ref{fig:adisr}). Therefore, large dust aggregates are unlikely to survive in the atmosphere unless the RAT efficiency of dust aggregates is substantially reduced compared to compact and composite grains. The disruption by RATD reduces the abundance of large grains and allow the distribution of dust over a large altitude because the sediment by gravity is less efficient for small grains. This RATD effect can thus explain the presence of high-altitude clouds in hot Jupiter and super-puff exoplanets. 



\subsection{Alignment of grains in the Earth atmosphere}
Although the present study is focused on hot Jupiter atmospheres, 
we here discuss the implication for grain alignment in the Earth atmosphere.
Observations of background starlight through dust clouds lifted off the Saharan desert reported the linear polarization \citep{2007ACP.....7.6161U}, and the authors suggested that grains are aligned along the vertical direction with the electric field in the Earth atmosphere. 
 Recent study by \cite{rm3} also study the alignment of grains in Earth atmosphere. However, both these studies disregard the effect of grain randomization by gas collisions. 
The altitude of Saharian dust clouds is estimated to be 3000-4000 m \citep{2007ACP.....7.6161U}. At such a height, the gas mass density of the Earth atmosphere is huge of $\rho\sim 10^{-3}\g\cm^{-3}$, corresponding to the molecular gas density $n_{\rm gas}\sim 2\times 10^{19}\cm^{-3}$, gas collisions are so efficient that can quickly damp the grain rotation in a time of
$\tau_{\rm gas}\sim 7.8\times 10^{-3}a_{-5}$ s (see Eq. \ref{eq:taugas}). 

In the RAT paradigm, grains can be aligned by RATs by sunlight as well as scattered light from the Moon or infrared radiation from the Earth. The mean wavelength of IR radiation by the Earth at $T_{\rm Earth}\sim 300\K$ is $\bar{\lambda}=0.53~\cm\K/T_{\rm Earth}=20\mum$ (see Eq. \ref{eq:wavemean_star}). 
With a high gas density in the Saharan dust cloud, even solar radiation cannot align grains. Therefore, an efficient alignment of Saharan dust is unlikely achieved by RATs. Nevertheless, the RAT alignment of grains at high altitudes is expected from Equation (\ref{eq:aalign_ana}) when the gas density decreases significantly to $n_{\rm gas}<10^{12}\cm^{-3}$ (see Figure \ref{fig:align}, the green light).

\section{Summary}\label{sec:summary}
We study grain alignment and rotational disruption of dust grains by RATs in hot Jupiter atmospheres. Our results are summarized as follows:

\begin{enumerate}

\item  We study the alignment of grains by RATs and found that small grains can be aligned. The minimum size of aligned grains decreases rapidly with altitude due to the decrease of the gas density.

\item We compare the different times of grain alignment with radiation, B-field, and E-field. We find that grain alignment occurs along the magnetic field due to a strong magnetic field in hot Jupiters for silicate grains or grain with iron inclusions, while carbonaceous grains can align with k-RAT.

\item Grain alignment with the magnetic field induces the polarization of starlight and polarized thermal dust emission, which paves the way for studying the magnetic fields of exoplanetary atmospheres with dust polarization.

\item We study the disruption of grains by the RATD mechanism and found that large grains can be disrupted rapidly. The disruption size decreases with altitude due to the rapid decrease of the gas density.

\item The disruption size by RATD decreases with decreasing the grain tensile strength. Compact grains are more easily to be disrupted than porous or fluffy grains of lower tensile strengths.

\item We suggest that the fragmentation of large grains into smaller sizes by the RATD effect reduces the gravitational sediment efficiency and facilitates the vertical transport of small dust to high altitudes. This may explain the presence of high-altitude clouds in hot Jupiter and super-puff atmospheres, as revealed by the flat transmission spectra observed toward hot Jupiters and super-puffs.

\end{enumerate}

\acknowledgments
T.H. acknowledges the support by the National Research Foundation of Korea (NRF) grants funded by the Korea government (MSIT) through the Mid-career Research Program (2019R1A2C1087045).

\bibliography{ms.bbl}
\appendix
\section{Planet temperature}
The most basic property of planets is its surface temperature. Assuming the black body for the planet surface, the temperature of the planet surface is obtained by the energy balance between heating and cooling
\bea
(1-A_{B})cu_{\rm rad}\pi R_{\rm p}^{2}&=&(1-A_{B})\left(\frac{L_{\star}}{4\pi D_{\rm p}^{2}}\right)\pi R_{\rm p}^{2}\nonumber\\
&=&\frac{4\pi}{f} R_{\rm p}^{2}\sigma T_{sf}^{4} ,
\ena
where $A_{B}$ is the Bond albedo, defined as the fraction of the light reflected by the atmosphere, $f$ is the factor characterizing the planet surface that the heat is distributed depending on the heat diffusivity through the atmosphere; $f=1$ for the case the heat is evenly distributed over the entire planet surface and $f=2$ for the case the heat is only distributed on the near-side hemisphere only, which yields
\bea
T_{sf}&=&\left(f(1-A_{B})\right)^{1/4}\left(\frac{L_{\star}}{16\pi \sigma D_{\rm p}^{2}}\right)^{1/4}=T_{\star}\left(f(1-A_{B})\right)^{1/4}\left(\frac{R_{\star}}{2D_{\rm p}}\right)^{1/2}\\
&\approx& 1251\left(f(1-A_{B})\right)^{1/4}\left(\frac{L_{\star}}{L_{\odot}}\right)^{1/4}\left(\frac{D_{\rm p}}{0.05\AU}\right)^{-1/2}\K.\label{eq:Tsf}
\ena

Dust grains cool rapidly due to IR emission. Therefore their temperature is lower, as 
\bea
T_{d}&\approx& 16.4U^{1/6}\K\nonumber\\
&=&861\left(\frac{L_{\star}}{L_{\odot}}\right)^{1/6}\left(\frac{D_{\rm p}}{0.05\AU}\right)^{-1/3}\K.
\ena 
Therefore, silicate dust grains can sublimate for hot Jupiters at $D_{\rm p}<0.01\AU$ where $T_{d}>T_{sub}\sim 1500\K$.

\section{Atmosphere model}
The equations of atmosphere are described by a set of differential equations for the fluid, including the continuity equation, mass conservation, and equation of state (see \citealt{Seager.2010}). Below, we review the fundamental equations for atmosphere.

\subsection{Isothermal model of atmosphere}\label{sec:apdx}
For the atmosphere in hydrodynamic equilibrium, the pressure variation with the height is described by
\bea
\frac{dp}{dr}=-\frac{GM_{p}\rho}{r^{2}}=-\frac{GM_{p}nm}{r^{2}},\label{eq:dpdr}
\ena
where $\rho=nm$ with $n$ the gas density.

The equation of state is
\bea
p=nkT_{\rm gas},\label{eq:eos}
\ena

Combining Equations (\ref{eq:dpdr}) and (\ref{eq:eos}), one obtains
\bea
\frac{dp}{p}=-\frac{GM_{p}mp}{kT_{\rm gas}}\frac{dr}{r^{2}},
\ena
where 

For simplicity, we assume the isothermal model for the atmosphere (i.e., $T_{\rm gas}=const$). Thus, the solution of the above equation provides the pressure decreases with the height as
\bea
p(r)=p(r_{0})\exp\left(-\frac{GmM_{p}}{kT_{\rm gas}rr_{0}}(r-r_{0})\right),\label{eq:pr}
\ena 
where $m=\mu m_{\rm H}$ is the mean particle mass with $\mu$ mean molecular weight (see also \citealt{Fortney:2005dl}).

The gas number density decreases with the radius $r$ as follows
\bea
n_{\H}(r)=n_{p}\exp\left[\frac{GM_{\rm p}}{v_{T}^{2}} \left(\frac{1}{r}-\frac{1}{R_{\rm p}}\right)\right],
\ena
where $v_{T}^{2}=kT_{\rm gas}/m$ is the thermal speed.

\subsection{$p-T$ model}
The isothermal model is not correct in realistic conditions. In reality, the temperature changes with the height due to various effects such as photoelectric heating and cooling, heat conduction and convection. The equations are similar to those in stellar interior.

\subsection{Roche limit}
The gravity effect is described by the Roche limit, the maximum radius where the gravity is still to dominate the gas. For a planet, the Roche limit is 
\bea
R_{\rm Roche}=\left(\frac{M_{\rm p}}{3M_{\star}}\right)^{1/3}D_{\rm p}=4.5R_{\rm p}\left(\frac{M_{\rm p}}{0.7M_{J}}\right)^{1/3}\left(\frac{M_{\odot}}{M_{\star}}\right)^{1/3}
\ena

\end{document}